\documentclass{ws-ijmpcs}

\usepackage{amsmath}
\usepackage{amssymb}

\begin{document}

\newcommand{\be}{\begin{equation}}
\newcommand{\ee}{\end{equation}}
\newcommand{\diff}{\ \mathrm{d}}
\newcommand{\sdiff}{\mathrm{d}}
\newcommand{\del}{\partial}
\newcommand{\unit}{\ \mathrm}
\newcommand{\unitb}{\mathrm}
\renewcommand{\vec}{\mathbf}

\markboth{S. Richter, F. Spanier}
{A spatially resolved SSC shock-in-jet model}

%
\catchline{}{}{}{}{}
%

\title{A SPATIALLY RESOLVED SSC SHOCK-IN-JET MODEL}

\author{Stephan Richter\footnote{srichter@astro.uni-wuerzburg.de}}

\address{ITPA, Universit\"at W\"urzburg, Emil-Fischer-Stra\ss e 31\\
W\"urzburg, 97074,
Germany}

\author{Felix Spanier\footnote{fspanier@astro.uni-wuerzburg.de}}

\address{ITPA, Universit\"at W\"urzburg, Emil-Fischer-Stra\ss e 31\\
W\"urzburg, 97074,
Germany}

\maketitle

\begin{history}
\received{Day Month Year}
\revised{Day Month Year}
\end{history}

\begin{abstract}
In this paper a spatially resolved, fully self-consistent SSC model is presented. The observable spectral energy distribution (SED) evolves entirely from a low energetic delta distribution of injected electrons by means of the implemented microphysics of the jet. These are in particular the properties of the shock and the ambient plasma, which can be varied along the jet axis. Hence a large variety of scenarios can be computed, e.g. the acceleration of particles via multiple shocks. Two acceleration processes, shock acceleration and stochastic acceleration, are taken into account. From the resulting electron distribution the SED is calculated taking into account synchrotron radiation, inverse Compton scattering (full cross section) and synchrotron self absorption. The model can explain SEDs where cooling processes are crucial. It can verify high variability results from acausal simulations and produce variability not only via injection of particles, but due to the presence of multiple shocks. Furthermore a fit of the data, obtained in the 2010 multi-frequency campaign of Mrk501, is presented.

\keywords{active galaxies; jets; Mrk501.}
\end{abstract}

\ccode{PACS numbers: 98.54.Cm, 98.62.Nx}

\section{Introduction}
Synchrotron Self Compton (SSC) models have been quite successful in explaining the emissions of blazars. However, recent observational results have led to the conclusion that the approach of modeling blobs in blazar jets as homogeneous regions employing SSC codes is not sufficient. Especially the observation of intra day variability introduces strong constraints for both the size of the emission region and the Doppler factor that usually lack observational evidence or even contradict those (e.g. Ref.~\refcite{mrk501_vlbi} and Ref.~\refcite{rapid_tev_var}). Furthermore the observed time delays between light curves of different frequencies, especially soft lags\cite{soft_xray_lag}, are hardly explainable by such models.

Two-zone-models like Ref.~\refcite{weidinger2} can produce scenarios that can reproduce observational data by merely varying physical parameters. Although this is an enormous improvement in comparison to one-zone-models, into which the electron distribution enters as a couple of unphysical parameters, there are still problems:

The stochastic character of the dominant acceleration process (Fermi-I) is not taken into account. It is modeled as a convection in momentum space and enters as a term in the Fokker-Planck equation. Furthermore the radiation zone is modeled homogeneously. Therefore variations, for example introduced via particle injection from the acceleration zone, will affect the entire space immediately. This behavior will lead to shortened timescales. Finally the cooling of the particles happens everywhere quasi locally and not depending on distance to the acceleration site.

Additional to addressing the aforementioned issues spatially resolved models can predict the morphology of the emission region. Using VLBI this is in principle observable.
\section{Model}	
The here presented model connects the dominant acceleration process with the geometry of the simulation box, resulting from the spatial discretization.

\subsection{Geometry}
\begin{figure}[ht]
\centerline{\psfig{file=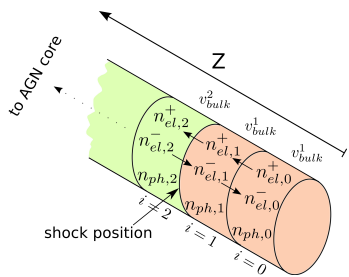,width=4.5cm}}
\vspace*{8pt}
\caption{A schematic illustration of the used geometry. The spatial discretization (index $i$) is along the jet axis. A shock is represented as a jump in velocity of the background plasma. \label{geometrie}}
\end{figure}

The discretization along the z-coordinate, e.g. along the jet axis, is shown schematically in Fig.~\ref{geometrie}.
The resulting slices are indexed with $i$. For each of them a bulk velocity is calculated depending on the positions and properties of the shocks, namely the compression ratio $R$ and shock speed $V_S$. The downstream velocity behind each shock $V_P$, expressed in the upstream frame is
\be
  V_P=\frac{V_S(R-1)}{R-V_S^2}\quad.
\ee
The Fermi-I acceleration then evolves naturally if particles undergo convection along $z$ as well as pitch angle scattering. In order to obtain a modeling for this process while keeping the momentum space one dimensional, we integrate over the pitch angle in each half-space separately. Hence we obtain two particle distributions - $n_i^+$ for particles moving downstream and $n_i^-$ for particles moving the opposite direction. The pitch angle diffusion is then modeled as scattering from one half-space into the other by a rate that enters the model as a parameter.
\subsection{Kinetic Equations}
To derive the kinetic equation all relevant processes have to be included into the Fokker-Planck equation, which is then discretized in $z$. After performing the integration as described above one obtains the convection equation in space as well as the decoupled kinetic equation
\be
  \label{eq:gamma_kinetic_el}
  \frac{\del n(\gamma)}{\del t}=\frac{\del}{\del \gamma}\left[D\gamma^2\cdot\frac{\del n(\gamma)}{\del\gamma}+(\beta_s\gamma^2-2D\gamma+P_{IC}(\gamma))\cdot n(\gamma)\right]+S(z,\gamma,t)\quad,
\ee
that describes the time evolution of the electron distributions $n_i^{+/-}$, i.e. for each spatial slice and half-space. In Eq.~\ref{eq:gamma_kinetic_el} acceleration due to Fermi-II is described via the momentum diffusion coefficient $D$. Furthermore losses due to synchrotron emission ($\beta_s$) and inverse compton scattering ($P_{IC}$) are included. $S$ stands for the source function.

To obtain the SED in each slice, the photon density is calculated employing
\be
  \label{kinetic_ph}
  \frac{\del N}{\del t}=-c\cdot\kappa_{\nu,SSA}\cdot N+\frac{4\pi}{h\nu}\cdot(\epsilon_{\nu,IC}+\epsilon_{\nu,sync})-\frac{N}{t_{esc}}\quad.
\ee
The dominant gain $\epsilon_{\nu,sync}$ is due to synchrotron emission and calculated employing the Melrose-Approximation\cite{brown}. The synchrotron self absorption coefficient $\kappa_{\nu,SSA}$ is calculated with the delta approximation. To obtain the rates for inverse Compton scattering $\epsilon_{\nu,IC}$ the full Klein-Nishina cross section\cite{blumenthal_gould} is used. The catastrophic loss is parameterized by the escape timescale $t_{esc}$. The full definitions of all used terms can be found in e.g. Ref.~\refcite{weidinger1}. Finally the total SED is calculated using the model of Blandford and K\"onigl\cite{blandford_konigl}, taking into account light travel times and time dilation.
\section{Results}
\begin{figure}[ht]
\centerline{\psfig{file=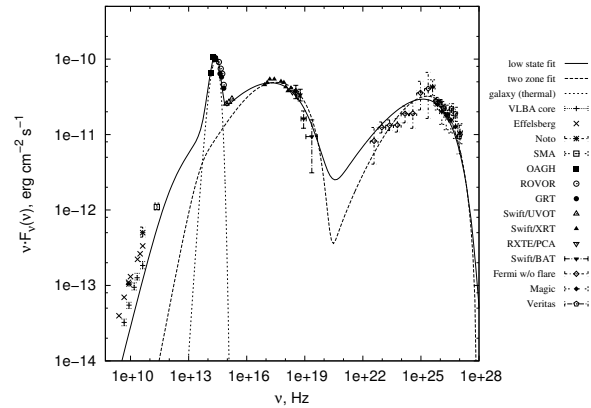,width=8cm}}
\vspace*{8pt}
\caption{Fit of the data obtained during a multi-frequency campaign. The comparison to the two-zone model fit, provided by Ref.~\protect\refcite{weidinger2}, shows much better agreement, especially in the radio regime.}
\label{mrk501_low}
\end{figure}

 The Blazar \textit{Mrk501}, a HBL (high frequency peaked BL-Lac object) with a redshift of $z=0.034$, was observed in a multi-frequency campaign between March and August 2009\cite{mrk501_campaign}. Our fit of the averaged low state of this source is shown in Fig.~\ref{mrk501_low}.
This fit was realized in a simulation region of $10^{15}\unit{cm}$ with a magnetic background field of $1\unit G$. Particles were injected with a rate of $\sim200 \unit{cm}^{-3}\unit s^{-1}$ and a Lorentz factor of $\gamma_{inj}=345$ far upstream i.e. at the edge of the simulated region. The shock responsible for the acceleration had a velocity $v_S=0.2\unit c$ and a compression ratio $R=3$. The Doppler factor of the shock frame is $\delta=20$. Although the size of the simulation region is not resolvable in observations, analysis of the produced morphology show inhomogeneities that affect the overall SED.
\section{Conclusion}
With the here presented model it is possible to confirm results regarding the particle acceleration from one- and two-zone-models and therefore trace them back to the jets microphysics. The number of parameters is not higher than that of homogenous SSC models and could be further reduced by connecting momentum diffusion and the scattering parameter, since both processes occur due to the presence of Alfv\'en waves. In the fit of the Mrk501 data the difference to homogeneous models is most prominent in the radio range. This effect is presumably due to the emission of electrons far away from the shock that are already cooled.

Furthermore our model is able to explain very short time variability since the lower limit for the timescale, due to the light crossing time of the emission region, no longer holds. Fitting of actual lightcurves will be subject to future publications.


\end{document}